# Molecular dynamics simulation of UO$_2$ nanocrystals melting under isolated and periodic boundary conditions

A.S. Boyarchenkov[a], S.I. Potashnikov[a], K.A. Nekrasov[a], A.Ya. Kupryazhkin[a]

[a] Ural Federal University, 620002, Mira street 19, Yekaterinburg, Russia

boyarchenkov@gmail.com  potashnikov@gmail.com  kirillnkr@mail.ru  kupr@dpt.ustu.ru

**Abstract**

Melting of uranium dioxide (UO$_2$) nanocrystals has been studied by molecular dynamics (MD) simulation. Ten recent and widely used sets of pair potentials were assessed in the rigid ion approximation. Both isolated (in vacuum) and periodic boundary conditions (PBC) were explored. Using barostat under PBC the pressure dependences of melting point were obtained. These curves intersected zero near –20 GPa, saturated near 25 GPa and increased nonlinearly in between. Using simulation of surface under isolated boundary conditions (IBC) recommended melting temperature and density jump were successfully reproduced. However, the heat of fusion is still underestimated. These melting characteristics were calculated for nanocrystals of cubic shape in the range of 768–49 152 particles (volume range of 10–1000 nm$^3$). The obtained reciprocal size dependences decreased nonlinearly. Linear and parabolic extrapolations to macroscopic values are considered. The parabolic one is found to be better suited for analysis of the data on temperature and heat of melting.

## 1. Introduction

Melting point of uranium dioxide (UO$_2$) is one of the most important characteristics in terms of safe management of nuclear fuel. In the event of a reactor malfunction the temperature at the center of a fuel rod can increase, leading to its melting. In the reactor, UO$_2$ fuel is exposed to high temperatures and radiation, and these conditions vary greatly across the fuel element (in connection with the temperature gradient, high burn-up [1]) and depending on the load time (accumulation of decay products and increase of oxygen partial pressure). As a result, the melting temperature depends on many factors. In particular, melting point and thermal conductivity are reduced in hyperstoichiometric UO$_2$ [2] or in the presence of transuranium elements (this dependence should be systematically investigated for the development of the fuel cycle in the fast breeder reactors). Unfortunately, experimental measurements are hampered by high temperatures (~3000 K), pressures (~100–1000 MPa) and radioactive toxicity, but molecular dynamics (MD) simulation is free from these constraints and allows one to estimate the melting point of UO$_2$ in the aforementioned conditions.

The publications dedicated to MD simulation of uranium dioxide melting have appeared only since 2008. In particular, our research group has simulated the melting of nanocrystals with surface, which were surrounded by vacuum under isolated boundary conditions (IBC), see [3]. However, that study used rather old interatomic potentials of Walker and Catlow [4], which were obtained in the harmonic approximation from the elastic properties at zero temperature. These potentials poorly reproduce the recent data on high-temperature thermophysical properties (thermal expansion, specific heat capacity, etc.) and characteristics of diffusion (formation and migration enthalpies of point defects), so they give inadequate characteristics of phase transitions (see [5] [6] for details). In addition, our old works were constrained by size of the crystals (no more than 10 000 particles) and simulation time (no more than 1 ns).

Govers et al. have conducted a comparison of sets of pair potentials (SPP), where the equilibrium temperature of two coexisted phases was determined by MD simulation of system consisting of both solid and liquid parts, for 5 SPPs [7]. The calculations were performed under periodic boundary conditions (PBC) on a system of 3000 ions (10×5×5 FCC unit cells) with 200 ps for each numerical experiment. The overestimation of melting point by 300–900 K for the four SPPs out of five was explained by the fact that no electronic defects were simulated.

A recent article by Arima et al. [8] is entirely devoted to MD simulation of melting under PBC. A comparison of the characteristics of single-phase (of 324 ions) and two-phase (of 648 ions) systems was made using three SPPs, and, in addition, the dependence of melting point on the concentration of Schottky defects was measured for potentials of Yakub et al. [9]. However, the melting temperature decreased only by ~50 K at 8% of the defects.

As it can be seen, the previous attempts of MD simulation of uranium dioxide melting were constrained by rather small sizes and times of several nanometers and nanoseconds (a few thousands of particles and a few hundreds of thousands of MD steps), while modern natural experiments, by contrast, deal with larger sizes of at least 10–100 nm (~0.1–1 million particles) and times of ~100–1000 ns. Use of the graphics processing units (GPU) as high-performance systems of parallel computing (e.g., NVIDIA CUDA technology) has allowed us to develop a fast parallel molecular dynamics program [10] [11] [12], which makes the direct comparison of the model with experiments possible.

Besides, we believe that the presence of surface or Schottky defects should significantly lower the melting point compared with the results of defect-free crystal simulation under PBC. Therefore, in this paper we study the melting of $UO_2$ cubic nanocrystals (CNC) with surface (surrounded by vacuum under IBC), volumes up to 1000 nm$^3$ (50 000 particles) and simulation times up to 100 ns (20 million MD steps) using 10 recent and widely used SPPs in the rigid ion approximation.

## 2. Methodology

The model of this work is isolated nanocrystal of uranium dioxide surrounded by vacuum (i.e., under IBC), which is formed of N rigid ions using the face centered cubic (FCC) unit cell with zero dipole moment. The interaction between particles is described by pair potentials composed of the long-range Coulomb term with the short-range Buckingham and Morse terms:

$$U_{ij} = \frac{Q^2 K_e q_i q_j}{r_{ij}} + X_{ij} \exp(-Y_{ij} r_{ij}) - \frac{Z_{ij}}{r_{ij}^6} + \\ + G_{ij}\left((\exp(-H_{ij}(D_{ij} - r_{ij})) - 1)^2 - 1\right) \quad (1)$$

Here $U_{ij}$ is the pair potential with partial charges defined by ionicity Q and the short-range parameters X, Y, Z, G, H, D (which are given in review [6] for the 10 considered SPPs); $K_e$ = 1.4399644 eV·nm is the electrostatic constant.

In this study, as in the previous one [6], we examined 10 recent and widely used empirical SPPs for $UO_2$ in the rigid ion approximation: Nekrasov-08 [3], Walker-81 [4], Busker-02 [13], Goel-08 [14], obtained in the harmonic approximation from elastic properties at zero temperature; Morelon-03 [15], obtained using the lattice statics from energy of Frenkel and Schottky defects; Yamada-00 [16], Basak-03 [17], Arima-05 [18], MOX-07 [5] and Yakub-09 [9], obtained using MD simulation of thermal expansion and bulk modulus.

The resulting forces and system energy were calculated as superposition of N(N–1)/2 independent pair interactions. In order to integrate the Newton equations of motion we used the well-known Verlet method [19] with a time step of 5 fs, correction of displacement of the center of mass and rotation around it, as well as quasicanonical dissipative Berendsen thermostat [20] with a relaxation time of 10 ps.

At each step of the dynamics, we calculated the instantaneous numerical density of particles n(t) in the system by averaging over an ensemble of S spherical layers using ion count and volume of each layer. Then, we calculated an average lattice parameter <a> from the time-averaged density and the constant ion count per FCC unit cell:

$$n(t) = \frac{1}{S}\sum_{s=1}^{S} \frac{N_s(t)}{V_s(t)}; \quad \langle a \rangle = \left(\frac{12}{\langle n(t) \rangle}\right)^{1/3} \quad (2)$$

In order to detect the melting in CNC during MD simulation we observed the self-diffusion coefficient of uranium cations, which in the liquid phase increases by several orders of magnitude and becomes close to the diffusion coefficient of oxygen anions, and also the jumps of density and energy of the system. We measured the melting temperature ($T_{melt}$) with a step of 10 K and a simulation time of 10 ns (2 million MD steps), i.e. we looked for a temperature value at which the original CNC melts during this time interval, while at 10 K less it does not melt. Temperature for SPPs Goel-08, Yakub-09 and MOX-07 were measured with a step of 5 K, which allowed slightly better fitting of size dependences.

Also we simulated melting under PBC using NPT ensemble, Ewald summation and Berendsen barostat (which is helpful in obtaining pressure dependences). In this case we conducted a series of simulations at temperatures where melting occurs with a step of 1 K and simulation time of 500 ps ($10^5$ MD steps). After that, from the whole series of length 400 K (or more if needed) we chose a temperature interval of length 10 K, which included at least three melting events (detected by sharp changes in density and enthalpy). Determining the melting point by one such event would be less reliable due to a stochastic kinetic initiation of this phase transition. Fig. 1 shows that uncertainty interval has a length of 100–200 K and too short simulation time makes it difficult to distinguish between solid and melted states (see "25 ps" curve). On the other hand, use of sufficiently long time causes slight shift to lower temperatures and the phases become clearly distinguishable (see "500 ps" curve).

Such thorough technique is very costly when applied to large isolated nanocrystals. Fortunately, they have uncertainty interval of only 30 K instead of 100 K (see Fig. 2), that is why using step of 10 K gives acceptable precision. Besides, it is possible to use the binary search algorithm in order to further reduce amount of computations.

## 3. Melting under PBC

We carried out MD simulations with different system sizes (up to 49 152 ions). Table 1 shows that the melting points obtained through MD simulations under PBC are weakly dependent on system size: for most SPPs the values differ by less than 50 K, starting with a system of 324 ions. However, such thermophysical quantities as energy (enthalpy), density, self-diffusion coefficients and their derivatives saturate at 1500 ions [6], that is why PBC results in the charts and tables are given only for this size.

The most recent and complete (more than twenty sources since 1953) review of experimental measurements of uranium dioxide melting point is presented by Baichi et al. in [21]. They recommended the value of Latta and Fryxell (3138 ± 15) K [22], which is also compatible with the data of liquidus and solidus of nonstoichiometric uranium dioxide. The largest presented there is a value of (3225 ± 15) K obtained by Ronchi et al. [23]. In the most recent experiments of Manara et al. [2], which, in particular, take into account the influence of rapid non-congruent evaporation, a value of (3147 ± 20) K was obtained.

Table 1 shows that MD simulations under PBC overestimate the experimental melting point of (3147 ± 20) K [2] by more than 600 K, as under PBC the crystals melt in a superheated state due to lack of surface (or other embedded defects of cationic sublattice). In order to bypass this effect retaining PBC, some authors [7] [8] measured the temperature of two-phase equilibrium crystal-melt (i.e., binodal). However, in such a case, the system volume would be controlled by barostat according to the melted half (as in the case of $UO_2$ it has a lower equilibrium density), and this would set up negative pressure in the solid half, which lowers the melting point according both to the Clausius-Clapeyron relation and our simulation results. Contact of crystal and melt additionally lowers the measurable melting point (which is also confirmed by our simulations of isolated crystals with potentials Walker-81, where the melted surface lowers bulk melting temperature, see Section 4). Finally, in [8] the authors note that crystallization time was considerably higher than melting time, i.e. the chosen observation interval of 200 ps could be too short.

As a result, the melting temperatures obtained for the two-phase system are lower than values for one-phase system. But although they were closer to the experimental estimates, such lowering has been achieved under the conditions that are farther from the conditions of nature experiments, that is why one cannot rely on them.

Therefore, we believe that it is more correct to investigate the melting of nanoscopic crystals with surface (which are isolated in a vacuum, i.e. with isolated boundary conditions – IBC).

However, Arima et al. [8] measured the melting point not only for the two-phase system, but also for the single-phase, so their results can be directly compared with ours. For the potentials Arima-05 melting temperatures coincide, for Basak-03 SPP our value turns out to be lower by 25 K, while for Yakub-09 SPP our value is greater by 135 K.

In the aforementioned paper [8] authors suggested a hypothesis of direct relationship between the melting temperature and the ionization coefficient Q, but it was tested only on three SPPs. However, as it can be seen from our results in Table 1 and the review of Govers et al. [7], this is not true (i.e., there is no simple relation). One can obtain large $T_{melt}$ for low Q, e.g. with Yamada-00 SPP, and small $T_{melt}$ for high Q, e.g. with Goel-08 SPP. Besides, in [7] the potentials of Karakasidis and Lindan [24] with formal (integer) ion charges had the lowest temperature of two-phase equilibrium. Also as one can see from the results of our previous work [6] there is no direct relation between Q and superionic transition temperature.

In addition, we also measured the melting temperature under PBC with artificial Schottky defects (trivacancies). With defect concentration of 0.8% (4 out of 500 molecules) the melting temperature is reduced by 200–300 K for Basak-03, Morelon-03, MOX-07, Yakub-09 SPPs. This is not consistent with the results of Arima et al. [8], where the melting temperature decreased by only 60 K with embedding of 8% Schottky defects.

In order to clarify the phase boundary between solid and liquid $UO_2$ Arima et al. [8] measured the pressure dependence of melting temperature (in the two-phase system) for Yakub-09 potentials. The authors obtained a linear dependence in the range from 0 to 3 GPa, the slope of which was a little higher than the experimental slope [2] for the range of 0–0.25 GPa. This thermodynamic dependence is described by the Clausius-Clapeyron equation in the form:

$$\frac{dT_{melt}}{dP} = \frac{T_{melt}\Delta V}{\Delta H}, \quad (3)$$

where $\Delta V$ is difference between the molar volumes of crystal and melt, $\Delta H$ is enthalpy difference (heat of fusion). Substituting our results into this equation we can verify correspondence of MD-simulations in the approximation of pair potentials and rigid ions with principles of thermodynamics.

We measured the melting temperatures (see Fig. 3) under positive pressures up to 25 GPa, where melting curves reach saturation, and under negative pressures down to –25 GPa, where all of them reach zero (i.e., the crystal lattice is unstable at all temperatures). Be advised that negative pressure is not purely theoretical concept, but occurs in nature and is achievable in the experiments [25], for example, upon wave unloading after shock compression. It is shown experimentally in [26] that dependences obtained under external pressure smoothly extend into the isotropically stretched state without any singularity when passing P = 0.

Fig. 3 shows that the linear form of the Clausius-Clapeyron equation used in works [2] [8], is approximate and applicable only at low pressures (e.g., at 3 GPa the deviation from line is 50–100 K).

As it can be seen from Fig. 3, the melting point of Busker-00 SPP is too high compared to the others over the entire range of pressures, with Walker-81 and Nekrasov-08 it is higher for positive pressures, with Yamada-00 − for negative. Interestingly, the potentials Yakub-09 and MOX-07, which reproduce other experimental data better than the rest [6], almost coincide on this figure over entire range of pressures.

The instability of crystal lattice, discovered for the potentials Yamada-00 at zero pressure [6], manifested itself at some negative pressures with potentials Walker-81, Busker-02 and Goel-08. Moreover, these four SPPs also have extreme intersections with axis $T_{melt}(P) = 0$: Yamada-00 and Busker-02 curves intersect it near –25 GPa, whereas Walker-81 and Goel-08 curves intersect it near –13 GPa.

For non-linear parameterization of the dependence $T_{melt}(P)$ the equation of Simon-Glatzel [27] is often used:

$$T_{melt}(P) = T_0\left(1 + P/P_0\right)^c, \qquad (4)$$

where $T_0$ is the reference melting temperature (e.g., under atmospheric pressure); $P_0 > 0$ and $c > 1$ are empirical parameters. This equation has an asymptote at negative pressures, the existence of which was confirmed experimentally [28].

In [29] the Simon-Glatzel equation was reduced to dimensionless form:

$$\frac{P'}{P_0} = \left(\frac{T_{melt}}{T_0}\right)^c, \qquad (5)$$

where $P = -P_0$ corresponds to zero melting temperature (the existence limit of a solid under isotropic stretching). Use of such dimensionless quantities makes it easy to compare the shape of melting curves. Thus, Fig. 4 shows that the curves for all SPPs can be divided into 2 groups (except for abnormal Yamada-00) – the first group consists of Basak-03, Yakub-09 and MOX-07, which reproduce temperature dependences of lattice constant and specific heat capacity better than the others [6], and the second group consists of all the rest. In addition, Arima-05 SPP is distinguished by the smoothest approaching to the horizontal axis $T = 0$, and Walker-81 SPP by the sharpest approaching to it.

In order to verify the Clausius-Clapeyron relation we fitted the dependences by quadratic function in the range from –10 to 10 GPa (taking into account the nonlinearity of the dependence $T_{melt}(P)$), and computed the derivative of this approximation at zero pressure (which is equal to the linear coefficient of this function). As a result (see Table 2), right-hand side of equation ($T_{melt}(P = 0)$ $\Delta V / \Delta H$) has turned out to be higher than the left ($dT_{melt}/dP$), and the difference between them increases sharply with increasing $dT_{melt}/dP$ up to 4 times (with Walker-81). One may notice that the melting point and the heat of fusion with different SPPs differ by no more than 2.3 times, and $\Delta V$ – by up to 12 times, so the values of $T_{melt} \Delta V / \Delta H$ differ mainly due to different $\Delta V$. Direct dependence of $dT_{melt}/dP$ on density jump also exists (and exceeds again the dependences on $T_{melt}$ and $\Delta H$), but it is weaker.

Our values $dT_{melt}/dP$ exceed the experimental estimates of 0.1 K/MPa [30] and 0.093 K/MPa [2], except for Yamada-00 SPP, which reproduce the experimental data generally worse, due to the lowest density jump of 2.5%. However, in a previous experimental study [33] the authors obtained a much larger value of ~0.2 K/MPa, which is closer to the results of our calculations. The best correspondence is shown by Basak-03, Yakub-09 and MOX-07 SPPs with values of 0.016–0.018 K/MPa. Table 2 shows that these potentials reproduce well the experimental density jump, therefore the main reasons for the deviation of model dT/dP values from the experimental estimates should be the higher (due to lack of surface) melting point and the lower heat of fusion. We can therefore expect a better correspondence for isolated nanocrystals with free surface.

Due to the fact that in molecular dynamics simulations under PBC the crystallization does not start within first 30 ps in the whole temperature range, we were able to draw the temperature dependence of relative density difference between crystalline and liquid $UO_2$ (see Fig. 5) and enthalpy difference between these two phases (see Fig. 6). The curves in Fig. 5 are increasing with temperature almost everywhere, which suggests that the density of the melt decreases with temperature faster than the density of the solid phase.

Density jumps at melting of potentials Basak-03 (8.4%), MOX-07 (9.1%) and Yakub-09 (10.3%) are closest to the experimental estimates of 7.3% and 9.6% (see Table 2), and curves for these SPPs in Fig. 5 converge to a single value of ~3% at low temperatures. With Yamada-00 SPP the densities of both phases differ by less than 3% everywhere, while in the interval $T < 3300$ K the density of melt is even higher than the density of crystal. Curves for all other potentials lie higher, especially Walker-81, for which the density jump is greater than 16% over the entire range of temperatures.

In the most recent experimental work on $UO_2$ melting [2] the density was not measured. However, the authors estimated the volume change $\Delta V$ using Clausius-Clapeyron equation. Their estimate was almost equal to the experimental value, but had lower uncertainty. Table 2 shows that Basak-03, MOX-07 and Yakub-09 SPPs overestimate those values by 20–50%, while other SPPs deviate stronger (up to 4 times).

As it can be seen from Fig. 6, the temperature dependence of enthalpy difference of crystalline and liquid $UO_2$ can be divided into 2 segments: at lower temperatures (less than 45% of $T_{melt}$ of the corresponding SPP) the dependence is practically constant, then "hills" can be observed, which roughly correspond to the superionic transition temperature. Enthalpy difference includes a heat of superionic transition prior to such a hill and after the hill it does not, so after the superionic transition the heat of fusion decreases monotonously.

For the majority of SPPs the heat of fusion (difference of the enthalpies at the corresponding melting point) falls in the range of 45–55 kJ/mol (see Table 2). Yamada-00 SPP has lower heat of 38.8 kJ/mol,

Busker-02 SPP has a 2-times greater heat of 90.8 kJ/mol, and only Arima-05 SPP with a value of 70.1 kJ/mol coincides with the experimental estimates of 70–78 kJ/mol [30] [34] [35] (which, given its very high melting point and density jump can be considered an accidental coincidence). The difference between the model and the experimental heat of fusion can be attributed to energy of defects (anionic, cationic or electronic, although the contribution of the latter defects is excluded in classical MD simulations). Therefore, there should be not a coincidence, but an understatement, which is observed for most SPPs.

Our values are lower than the values obtained for the two-phase system in [8]: ~60 kJ/mol for Basak-03 and Yakub-09 SPPs, ~105 kJ/mol for Arima-05. Probably it is due to lower concentration of defects at the corresponding melting temperatures, which are lower than our values in Table 1.

### 4. Melting point under IBC

The results of isolated crystals simulations are shown in Table 3, where one can see that at first the melting points of CNCs rise sharply with increasing size and then gradually reach saturation. Measurement of the CNC heat capacity showed that the temperature of superionic transition (anion sublattice disordering) was almost independent of CNC size and close to the values obtained for quasi-infinite crystals under PBC [6]. For example, it appeared to be lowered by ~100 K for the crystal of 2592 ions with MOX-07 SPP and saturates to the PBC value with size increase.

The superionic transition temperature in our simulations was not directly related to the melting temperature. So, four of the studied SPPs (Walker-81, Nekrasov-08, Arima-05, Goel-08) gave the melting points, which are lower than the corresponding temperatures of superionic transition measured in [6] for all sizes of CNC, while for the rest SPPs, with exception of Yamada-00 (which always melted at temperature above superionic transition), the size dependence of the melting point crossed the superionic transition temperature in the range of C=6–11 (where C is the number of FCC-cells on the edge of CNC).

With Walker-81 SPP, both cationic and anionic sublattices were disordered on the surface, but ordered in the bulk (i.e., superionic transition was also absent, but the surface was melted down). In addition, the smallest system (of 768 ions) was in amorphous state instead of melted. These conclusions are based on visual observations of the system during the numerical experiments and confirmed by analysis of the diffusion data (a smooth, without a jump, change of diffusion coefficients with temperature in the amorphous state and a large difference between the mobility of ions in the molten surface and ions in the bulk). As an example of the visual observations results, Figs. 7–8 show an illustrative projection of the uranium sublattice and a path of a central oxygen anion (in a time of 0.5 ns) for Walker-81 SPP (with solid bulk and liquid surface) and MOX-07 SPP (with surface and bulk in the superionic phase).

For comparison with experimental estimates obtained for macrocrystals it is necessary to extrapolate model data to a crystal of infinite size. A drastic reduction in a melting point of smaller crystals (by 800–1800 K compared to the largest crystal) also requires discussion.

In order to extrapolate the dependence to infinite size, we first plotted the dependence of melting point upon the quantity 1/C (see Fig. 9), where C is the number of FCC unit cells on edge of CNC (containing $N = 12 \cdot C^3$ ions correspondingly). Slopes of the curves for different SPPs turned out to be close to each other, with the exception of Walker SPP (due to the aforementioned disordered surface) and Yamada-00 SPP (which has shown instability of the anionic sublattice in the experiments under PBC [6]).

The left parts of these plots are almost linear, therefore extrapolation of them to intersection with the vertical axis (where $1/C = 0$) is an estimate of the macrocrystal melting point. The results of such extrapolation from the four largest sizes (25 000–50 000 ions) are shown in the row "Linear" of Table 3.

For more accurate estimate of the macrocrystal melting point we should take into account a non-linear behavior of the reciprocal size dependences in Fig. 9. Therefore, we also approximated these dependences by parabolas with their maximum lying on the vertical axis, which are described by the equation $T_{melt}(1/C^2) = T_0*(1-Y/C^2)$ and correspond to straight lines on the chart of the melting point versus squared reciprocal size (see Fig. 10). In this case the measured values fit a straight line much better, so we have built approximations from all the points available, making exception only in the case of the smallest sizes, inclusion of which led to a strong change of the slope and increase of the standard deviation. The greatest number of points (namely, four) excluded this way were for Yamada-00 SPP, the inadequacy of which in the temperature range 2000–2700 K were previously noted in different studies [6] [15] [17]. Results of the parabolic extrapolation are given in the row "Parabolic" of Table 3, where one can see that they are lower than the corresponding values of row "Linear" by 5–10%.

Fig. 10 shows these dependences in dimensionless form (with normalization to the extrapolated value of the melting point). It is seen that majority of potentials have almost the same slope of the curves, and melting point of the smallest crystals is lower by 20–25%. Whereas with Walker-81 and Yamada-00 SPPs it is lowered by 35%.

When the article was ready, we were asked by reviewer to assess the recent shell-core potentials Read-10 [36] regardless of rigid ion approximation in our

MD simulation code. We obtained parabolic extrapolation of (5225 ± 20) K with the slope of squared reciprocal size dependence close to the slope of Busker-02. This confirms high melting temperature of ~6600 K obtained with that SPP under PBC [6] and conclusion that shell-core potentials with formal charges are unsuitable for use without shells.

In a number of studies (e.g. [37] [38] [39]) the theoretical formula of Thomson was used in order to estimate the melting point of nanocrystals and to analyze the experimental data. We used its approximate (linear) version, as in [39]:

$$T_{melt}(R) = T_0 \cdot \left(1 - 2\sigma \cdot \frac{v}{q \cdot R}\right) \qquad (6)$$

Here R is the radius of surface curvature, $T_0$ is the macrocrystal melting temperature, $\sigma$ is the energy at the interface between solid and liquid phases; v is the volume per molecule; q is the latent heat of fusion (crystallization) per molecule. In fact, the ratio q/v is the heat of fusion per unit volume of the crystal.

In this study, extrapolating our data into the region of macrocrystals and comparing the predicted melting temperature to the experimental value, we have restricted ourselves to the use of the Thomson formula in the form:

$$T_{melt}(R_{eff}) = T_0 \cdot \left(1 - \frac{X}{R_{eff}}\right) \qquad (7)$$

At the same time, we tested the applicability of such (linear on the reciprocal size) approximation to small nanocrystals.

To do this we examined the dependence of the measured melting temperatures (see Table 3) of reciprocal effective radius $1/R_{eff}$ of model crystals, which was calculated using the formula $R_{eff} = (3V/4\pi)^{1/3}$ from their volume V, which was expressed via the lattice parameter (a) and the number of unit cells on the edge of the cube as $V = (aC)^3$.

The results of extrapolation by formula (7) are shown in Table 4. Since the lattice parameter varies with nanocrystal size not much (by less than 3%), the plots $T_{melt}(1/R_{eff})$ are similar to the plots $T_{melt}(1/C)$ of Fig. 9, so we omit them. Nevertheless, analysis by formulas (6) and (7), in contrast to the extrapolation in the coordinates $T_{melt}(1/C)$, allows us to estimate the interfacial energy $\sigma$ and verify that the values of parameter X, calculated from the largest sizes (25 000–50 000 ions), correspond to their physical meaning.

From formulas (6) and (7) we find that:
$$X = 2\sigma v/q \qquad (8)$$

If the estimates of the heat of fusion and the specific volume of uranium dioxide recommended in the review [30] q = 70 kJ/mol = 0.73 eV/molecule and v = 0.04703 nm$^3$/molecule are used, then for MOX-07 SPP the value of X = 0.345 nm corresponds to $\sigma$ = 2.68 eV/nm$^2$. There are no experimental data for $\sigma$, but its order of magnitude can be compared with the values of the surface energy density (or surface tension) of uranium dioxide in solid and molten states at the interface with vacuum. In [39] authors assumed that $\sigma$, which was the interfacial energy at the crystal-melt boundary for iron, is by the order of magnitude less than energies $\sigma_c$ (crystal-vacuum) and $\sigma_m$ (melt-vacuum). For uranium dioxide at melting point of 3150 K $\sigma_c$ = 0.45 J/m$^2$ ≈ 2.8 eV/nm$^2$ and $\sigma_m$ = 0.51 J/m$^2$ ≈ 3.2 eV/nm$^2$ [30]. So, the value of $\sigma$ given by formula (8) is indeed less than these quantities, but only by 5–20%.

The linearity of the squared reciprocal size dependence of melting point (Fig. 10) indicates that the slope of the reciprocal size dependence (Fig. 9) decreases with increasing system size. And if the quadratic formula holds true for systems of any size, then in the limit of infinite size this slope tends to zero. On the other hand, if the slope should be reduced not to zero, but by an order of magnitude, as taken in [39], then the parabolic model will still describe systems of small size well (like in this paper), but would require refinement for systems of over 50 000 ions. This is confirmed by thorough observation of the plots in Fig. 10, where the melting temperatures calculated for the largest CNCs are somewhat above the corresponding parabolic approximations. Therefore the value of $\sigma$ = 2.68 eV/nm$^2$ (obtained from the linear extrapolation (7) and experimental values of q and v) is just an upper bound, hence there is no contradiction with the physical meaning of the Thomson formula.

Only one SPP (Goel-08) has linear and parabolic estimates for macrocrystal melting point ((3051 ± 4) K and (2969 ± 2) K, respectively) that are lower than the experimental estimates: (3138 ± 15) K [21], (3147 ± 20) K [2] and (3225 ± 15) K [23]. Together with Goel-08, the best reproduction of the experiments is demonstrated by two SPPs: Yakub-09 (with estimates of (3223 ± 26) K and (3105 ± 3) K) and MOX-07 (with estimates of (3423 ± 1) K and (3291 ± 5) K).

The linear extrapolations (from the dependences $T_{melt}(1/C)$ or $T_{melt}(1/R_{eff})$) can be considered as upper bound of the melting point of infinitely large crystals, and the parabolic (from the dependences $T_{melt}(1/C^2)$ or $T_{melt}(1/R_{eff}^2)$) as the lower bound, which is also significantly more accurate (since the points for all sizes fit the straight line).

The dependence $T_{melt}(1/C^2) = T_0*(1-Y/C^2)$ could have a physical meaning and correspond to the Thomson formula (6) if the value of $\sigma$ for small crystals was not constant but proportional to $1/R_{eff}$. This assumption also explains the fact that the value of $\sigma$ we received for small crystals from Thomson formula (6) was greater than expected from experimental data (see discussion above).

The physical meaning of the depending on $1/C^2$ rather than 1/C could be due to the fact that the melting of nanocrystal starts from the surface, therefore it is determined by the surface area rather than the linear dimension.

The squared reciprocal size dependence of nanoparticle melting point has been also discussed by Farrell and Van Siclen in [40], but the authors explained this dependence by the covalent nature of chemical bonding in semiconductors.

Compared with the melting point measured under PBC, the parabolic extrapolations for nanocrystals are lower by almost 50% for Walker-81 SPP (due to the molten surface), by 26–34% for Busker-02, Goel-08 and Nekrasov-08 and by 20–23% (which is about 700 K) for all the rest. This substantial lowering emphasizes the important role of surface in simulation of uranium dioxide at high temperatures, particularly in the simulation of the melting process. It should be noted, that the melting temperature measured under PBC could be significantly lower (i.e., closer to IBC) if the system had surface in form of a cavity of several Schottky defects (trivacancies) as discussed in Section 3.

## 5. Heat of fusion

Although the heat of melting and the density jump for macroscopic crystals of uranium dioxide have been measured experimentally, their dependence on the system size remains unknown. Besides, in order to verify the quality of the model it is interesting to compare our values with the recommended experimental estimates.

First, we measured the equilibrium density and energy of the liquid phase of the systems of different sizes at the appropriate melting temperatures. To measure the density and energy of the solid phase at the same temperature, we relaxed the original CNC at a temperature lower by 10 K, and then after the equilibration raised the temperature to the desired value. This was done because initial crystals compared with equilibrium crystals have excess energy (which is gradually pumped out by the thermostat), so the latter melt at higher temperatures. Due to this excess energy the smallest CNC quickly melted, and subsequently crystallized again (see Fig. 12), however in another form. As it can be seen in Fig. 13, during the process of relaxation a shape of large nanocrystals also changed from cubic to a truncated octahedral (we have examined this process in detail in a separate article [41]). Because of the long duration of such numerical experiments we have conducted them only for six SPPs with the most appropriate melting point: Walker-81, Morelon-03, Basak-03, Goel-08, Yakub-09 and MOX-07.

Fig. 14 shows the plots of the specific heat of fusion versus squared reciprocal size. As in the analysis above, in this case the parabolic approximation (linear with regard to $1/C^2$) fits the measurements better than the linear ($1/C$). The results of such extrapolation are shown in Table 4. The data for the smallest system (of 768 ions) were excluded because their melting temperature is strongly underestimated, as such CNC in fact consists of the surface layer entirely.

It is seen that with our SPP MOX-07 the calculated heat of melting is close to a constant value of ~41 kJ/mol, the slope of its $q(1/C^2)$ dependence being 3–12 times less than with other potentials. Yet, the value of q somewhat increases with nanocrystal size, and its extrapolation to the macroscopic size makes up $(40.7 \pm 0.5)$ kJ/mol. The size dependence of the heat of fusion could be one of the reasons for nonlinearity of the plots of melting point versus reciprocal size (Fig. 9), but the slope of the heat of fusion dependence is lower by 1–2 orders of magnitude, i.e. it should not be the main reason of the nonlinearity. The corresponding parabolic extrapolation for Walker-81 SPP (31.8 kJ/mol) is again understated due to the molten surface, and extrapolations for the rest of SPPs give similar values in the range of 40–46 kJ/mol.

Compared with the value of $(70 \pm 4)$ kJ/mol recommended in the review [30], the results of MD simulation are underestimated by 25–30 kJ/mol. This discrepancy may be due to inclusion of the heat of superionic transition into the recommended heat of melting. In particular, the authors of that review note that their polynomial equation of enthalpy (which is the basis for the heat of fusion value) does not take superionic transition into account. However, this consideration does not apply to the potentials of Walker-81 and Goel-08 (which give the lowest heat of fusion), because with these potentials the CNC turns from the crystalline state directly into molten bypassing the superionic state, so the heat of fusion in the simulation takes into account the disordering of both anionic and cationic sublattices.

Another reason of this discrepancy may be due to partial inclusion of the heat of cationic Frenkel disordering into the recommended heat of melting. In the model crystals this type of disorder becomes essential within the last 100 K before fusion and probably becomes the mechanism of melting of the cation sublattice. On the other hand, the known experimental data on enthalpy and heat capacity of solid $UO_2$ do not reflect the appearance of the Frenkel disorder, which allows suggestion that in the experiment it has not been distinguished from melting. In the latter case the energy of Frenkel disorder becomes incorporated into the heat of fusion, increasing it relative to the model values.

Table 4 also shows that for all SPPs considered the heat of fusion obtained for quasi-infinite crystals (PBC) is greater than for the crystals with free surface (IBC), including extrapolations to the infinite size. This means that specific energy of the crystals with free surface at melting is closer to the liquid phase not only at the surface, but in the bulk also. So, under IBC the model crystals should be more disordered, which is true, because the surface allows formation of the Shottky defects and contributes to formation of the Frenkel pairs.

We also note that for estimation of σ from the Thomson formula (6), it would be correct to use the calculated values of q from Table 4 rather than the experimental value of the heat of fusion, since we consider the melting of the model crystals, not natural. In particular, with our potentials MOX-07 and the values of q = 40.7 kJ/mol, v = 0.04655 nm$^3$/molecule the value of σ is equal to 1.56 eV/nm$^2$ instead of 2.68 eV/nm$^2$. This value corresponds better to the fact that the surface energy at the interface of solid and liquid phases must be lower than the energy at the interface of solid phase and the vacuum.

The surface energy values close to values for our MOX-07 SPP are obtained for Yakub-09 (σ = 1.56 eV/nm$^2$ with v = 0.04655 nm$^3$/molecule) and Basak-03 (σ = 1.54 eV/nm$^2$ with v = 0.04649 nm$^3$/molecule) SPPs. Goel-08 gives lower value of surface energy (σ = 1.21 eV/nm$^2$ with v = 0.04694 nm$^3$/molecule), while higher values are obtained for Walker-81 (σ = 1.83 eV/nm$^2$ with v = 0.04164 nm$^3$/molecule) and Morelon-03 (σ = 1.73 eV/nm$^2$ with v = 0.04653 nm$^3$/molecule).

## 6. Density jump at melting

It is known that at the transition from solid phase to liquid the density of uranium dioxide decreases abruptly by 7–9% [30] (estimates of Ronchi and Christensen). The results of our calculations are shown in Fig. 15 and Table 4. It can be seen that our values of the density jump are linearly dependent on the effective reciprocal size $1/R_{eff}$. Extrapolation of this dependence to crystals of infinite size (see Table 4) for our potentials MOX-07 gives 7.97 ± 0.09%, which quantitatively coincides both with the value obtained under PBC (at the same temperature) and with the experimental values. We can therefore expect that our interatomic potentials reproduce well the density of the melt, at least near the melting point.

We excluded plots of Walker-81 and Goel-08 from Fig. 15, because their density jumps are in the ranges of 26–29% and 13–18% respectively, which could considerably decrease the detail level of this figure. Moreover, unlike more recent potentials, Walker-81 has almost constant density jump for crystals of over 6000 ions. The values for Basak-03 (7.47 ± 0.07%) and Yakub-09 (8.66 ± 0.12%) are within the tolerance of experimental data, while Morelon-03 with the value of (11.2 ± 0.1%) lies above the upper bound. It should be noted that for all the potentials the extrapolations of the density jump to macrocrystals are in very good agreement (differ by less than 0.1%) with values obtained under PBC at the same temperature.

The values of the right-hand side of Clausius-Clapeyron equation $T_{melt}\Delta V/\Delta H$ for isolated nanocrystals (see Table 4) are much closer to the values of the left-hand side obtained under PBC, except for the Walker-81 and Goel-08 with difference of 2–4 times, as their value of ΔV almost have not changed. For other SPPs decrease of ΔV by 25–40% due to IBC led to a better agreement with experimental and recommended values. Thus, Table 4 shows that with MOX-07 and Yakub-09 the deviation became less than 9%.

## 7. Conclusions

Use of the high-performance graphics processors (GPU) and NVIDIA CUDA technology in our parallel molecular dynamics (MD) program [10] [11] [12] has enabled us to conduct a simulation of melting of surrounded by vacuum (under isolated boundary conditions – IBC) cubic nanocrystals (CNCs) of uranium dioxide with volumes up to 1000 nm$^3$ (~50 000 particles) and simulation times up to 100 ns (20 million MD steps) using 10 sets of interatomic pair potentials.

For all the SPPs the melting points, measured on ideal lattice under periodic boundary conditions (PBC), are overestimated compared with the experimental estimates by more than 600 K due to the lack of surface or other sources of defects in this model. In particular, embedding of small cavity of 4 trivacancies (Schottky defect concentration of 0.8%) caused decreasing of melting temperature by 200–300 K.

For all the SPPs the obtained pressure dependence of melting temperature has a general form: it is monotonously increasing in the whole range of considered pressures, crosses zero at some negative pressure and saturates at ~25 GPa. Verification of the Clausius-Clapeyron thermodynamic relation showed that the differences between left and right parts of this relation are greater than in the work of Arima et al. [8]. The main reasons for deviations of MD simulations under PBC from the experiments are the overestimated melting point (due to lack of surface) and the underestimated heat of fusion (due to excluding the heat of superionic transition).

We obtained nanocrystal size dependencies of the melting temperature, the heat of fusion, as well as the density jump. The main result of the calculation is the conclusion that the melting temperature of uranium dioxide nanocrystals decreases substantially with decreasing size. And a large difference of melting temperatures between the simulations under IBC and PBC indicates the important role of surface at high temperatures.

The dependence of the melting temperature on the reciprocal size was nonlinear, therefore its analysis using the linear approximation of the Thomson formula (6) gave only rough upper bounds for macrocrystal melting point and energy at the interface of the two phases. In accordance with the Thomson formula, the increase in the slope of the curves with decreasing size can be associated with both a decrease in specific heat of fusion and with increase of the surface energy at the interface between solid and liquid phases. As the size decreases, the discrepancy between the linear approximation from the large CNCs and measured values of melting point reaches 400–900 K, hence for small CNCs the analysis requires taking into account the nonlinear terms of expansion (e.g., the second power of the reciprocal size).

For a more accurate extrapolation of the size dependences to the region of macrocrystals we used the parabolic approximation with zero linear term, which in the coordinates of $T_{melt}(1/C)$ corresponds to a parabola with a maximum at zero. Plots of the melting temperature and the heat of fusion versus squared reciprocal size $1/C^2$ were close to linear. The physical basis of such a parabolic extrapolation in our opinion is dependence of the melting process on crystal surface area (proportional to the square of its linear size). However, as shown, this approximation describes well the systems of up to 50 000 particles, and thus extrapolations into the region of macrocrystals could be underestimated by several tens of degrees. Therefore, in the perspective one should take into account the linear term of expansion with some small coefficient, which will require simulation of the systems of several hundred thousand particles.

Among the estimates of melting point the closest to modern experimental data (3147 K [2] and 3225 K [23]) were the values for Goel-08 (2969 K), Yakub-09 (3105 K) and MOX-07 (3291 K) SPPs. Overestimation of the experimental values by 7–10% is shown by Walker-81 (due to the molten surface), Morelon-03, and Basak-03 SPPs.

Estimates of the heat of fusion in this work were lower by 25–30 kJ/mol in comparison with the recommended value of 70 kJ/mol [30] for all sets of potentials, possibly due to the fact that our simulated values do not comprise the heats of anionic and cationic Frenkel disorder, which are included into the enthalpy of the solid phase instead.

Finally, the density jump at melting extrapolated in the coordinates (1/C) quantitatively coincides with the experimental estimates of 7–9% [30] [31] for Basak-03, Yakub-09 and MOX-07 potentials. The potentials Morelon-03 with a value of 11.2% give the value slightly above the experimental range. Walker-81 and Goel-08 potentials, which have been fitted using the simpler harmonic oscillators approximation and have not been able to reproduce the experimental curve of thermal expansion, yield highly overestimated values of 19.7% and 27.6%, respectively.

According to the results of this work, it should be noted that Yakub-09 and MOX-07 SPPs reproduce the characteristics of melting better than the potentials proposed previously. One can explain this advantage by the fact that in the process of their fitting high-temperature experimental data have been used more completely. In particular, the dependence of the thermal expansion on temperature in the entire temperature range including the regions of superionic transition and melting has been taken into account. Nevertheless, significant underestimation of the heat of fusion confirms a need of further development of pair potentials fitting methods (in particular, taking into account the experimental data for the liquid phase) and the model of particle interactions (e.g., consideration of electronic defects).

During the simulation of CNCs we revealed the phenomenon of structural relaxation accompanied by decreasing energy and the change in surface shape, presumably to a truncated octahedron. Part of the excess energy of the initial state is spent on crystal lattice disordering, which could lower the computed values of melting temperature and heat of fusion and affect their size dependence. Investigation of the structural relaxation process and comparison of model surface properties with known experimental data are discussed in our separate paper [41].

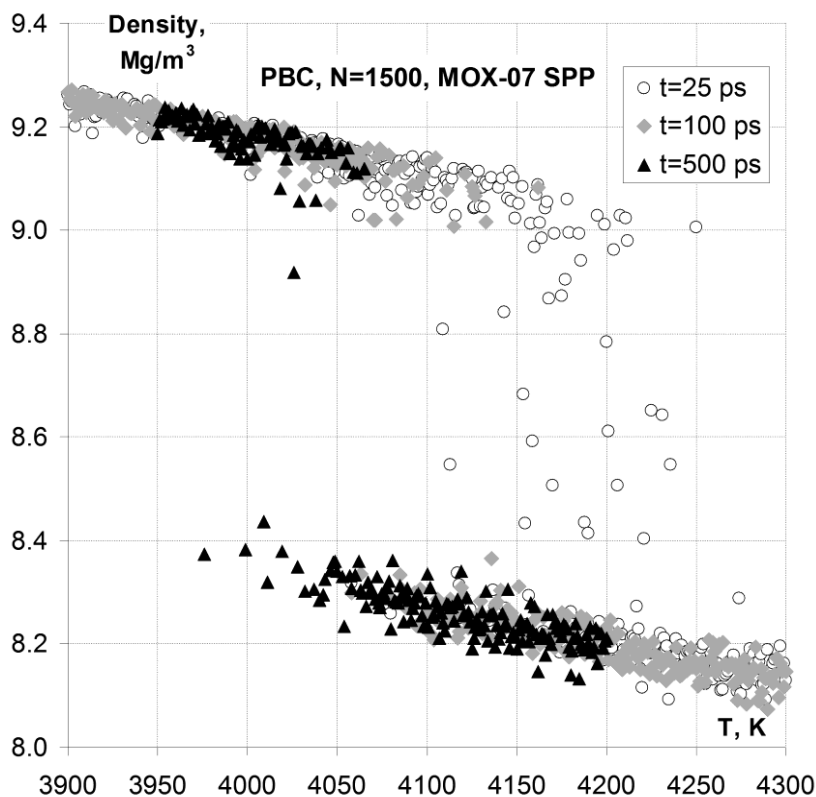

FIG. 1. Temperature dependence of density obtained under PBC with step of 1 K near melting point.

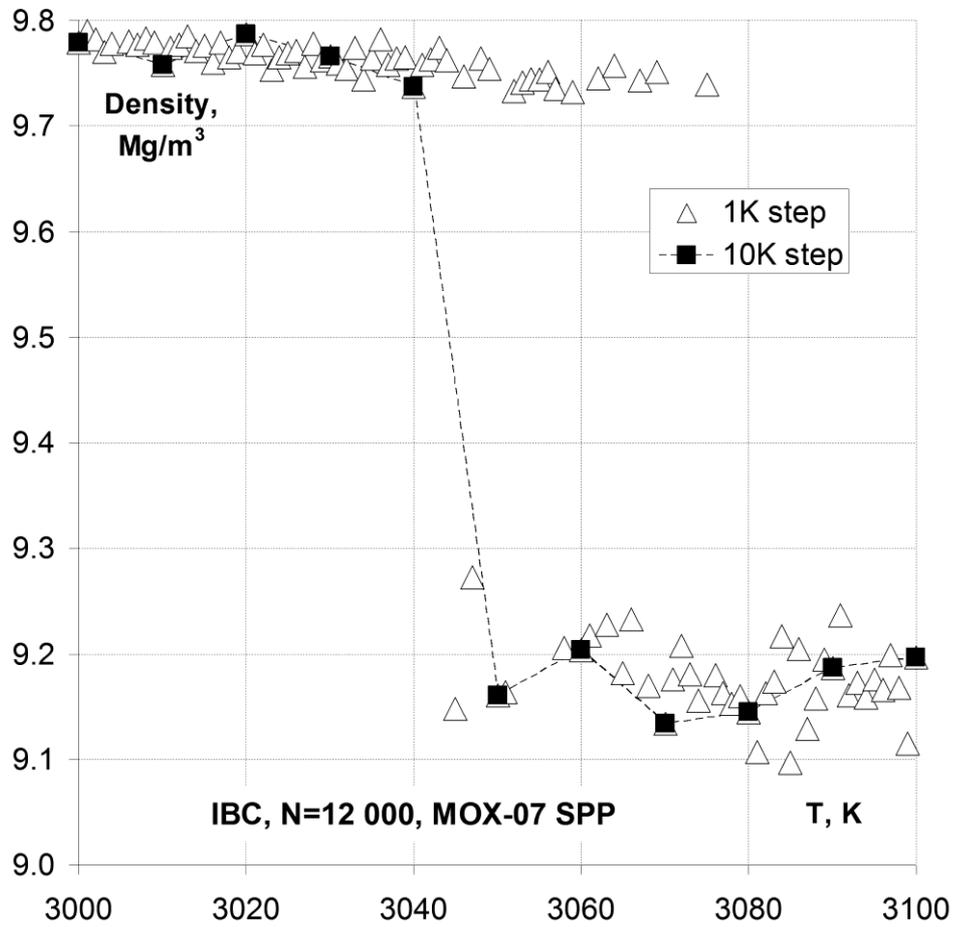

FIG. 2. Temperature dependence of density obtained under IBC with steps of 1 K and 10 K near melting point.

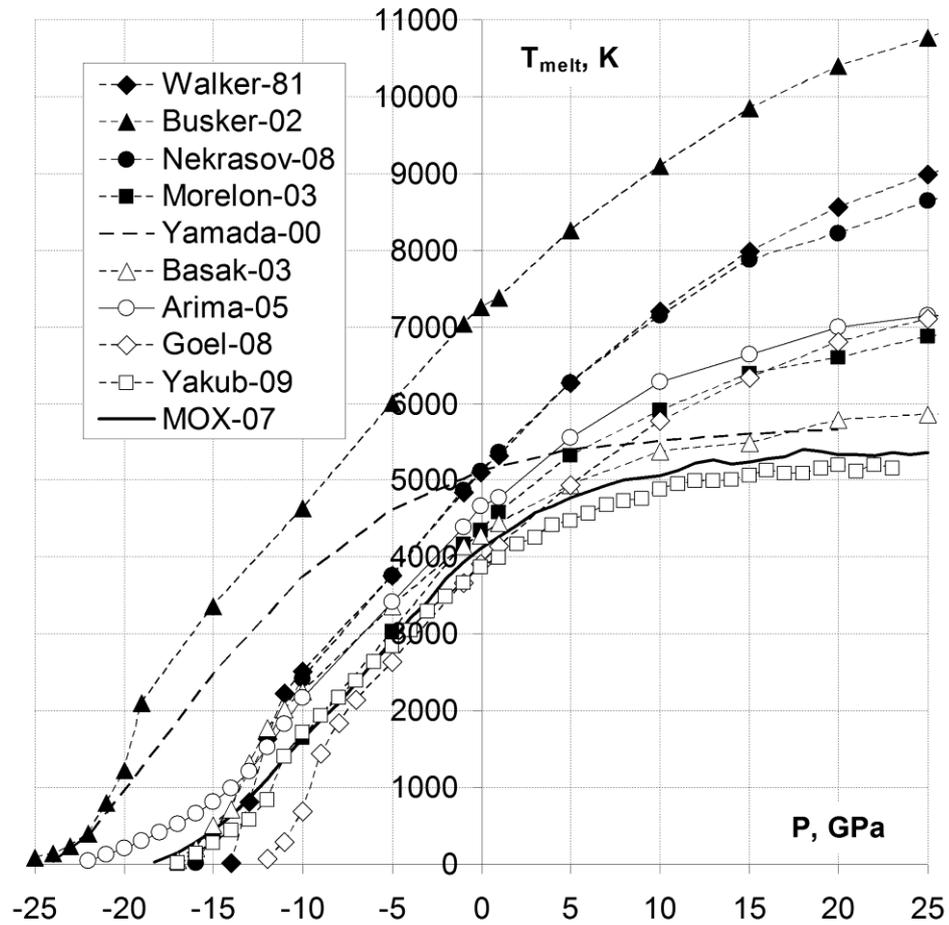

FIG. 3. Pressure dependence of melting temperatures under PBC.

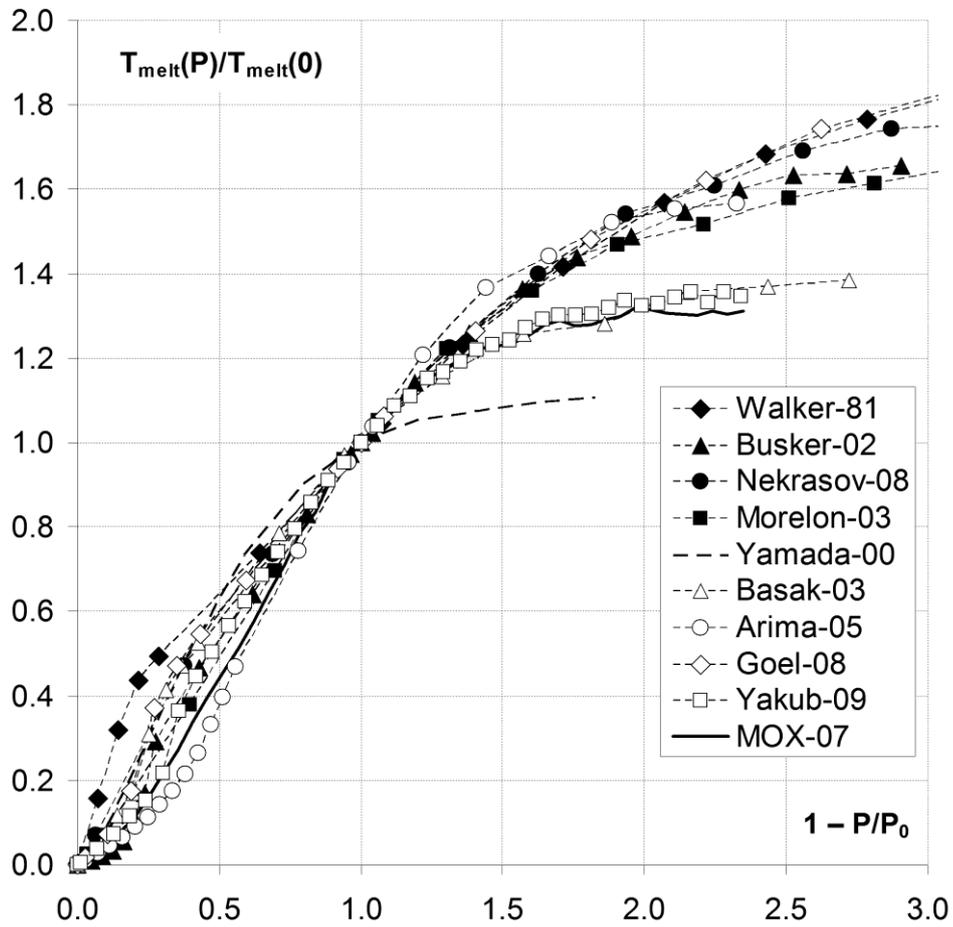

FIG. 4. Pressure dependence of melting temperatures under PBC in dimensionless form.

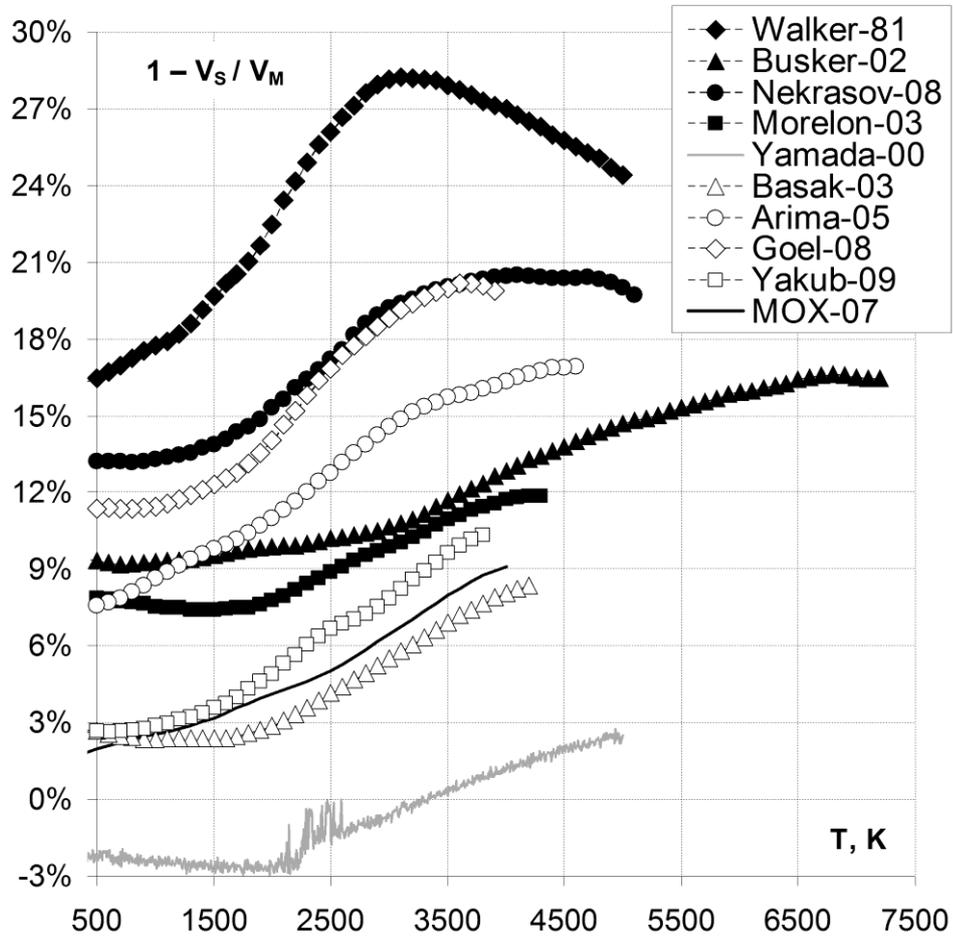

FIG. 5. Relative difference of densities between solid and liquid phases under PBC.

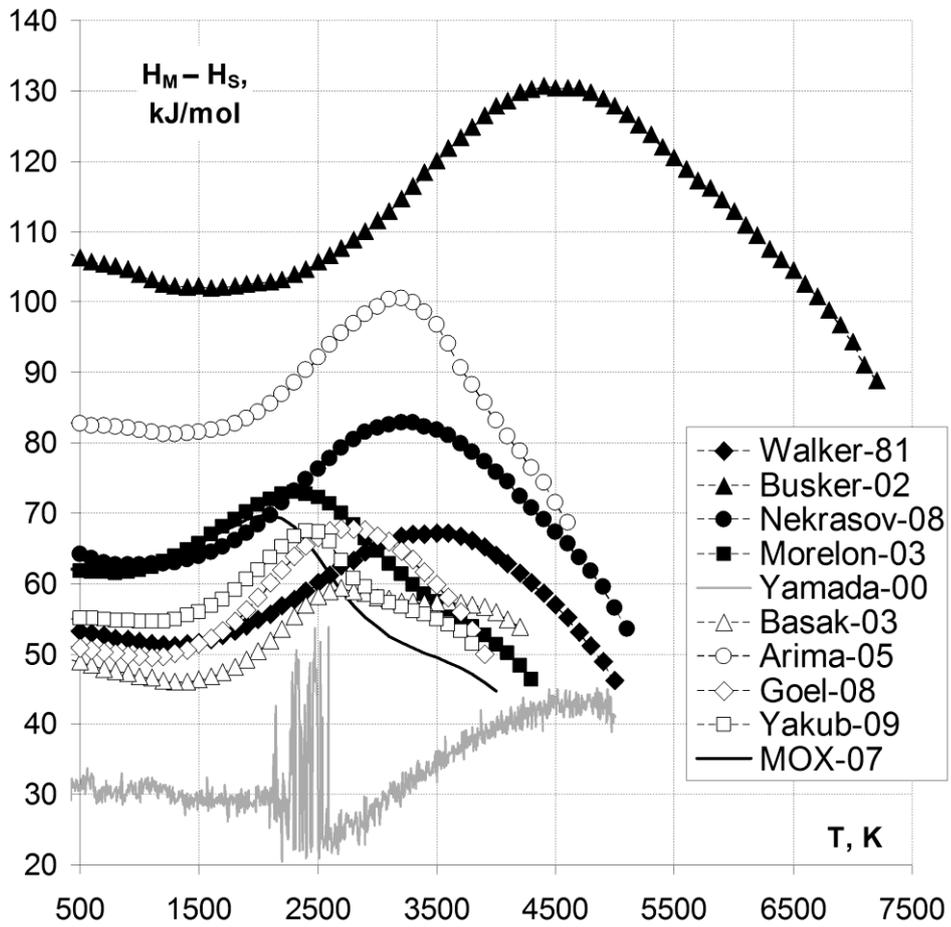

FIG. 6. Difference of enthalpies between solid and liquid phases under PBC.

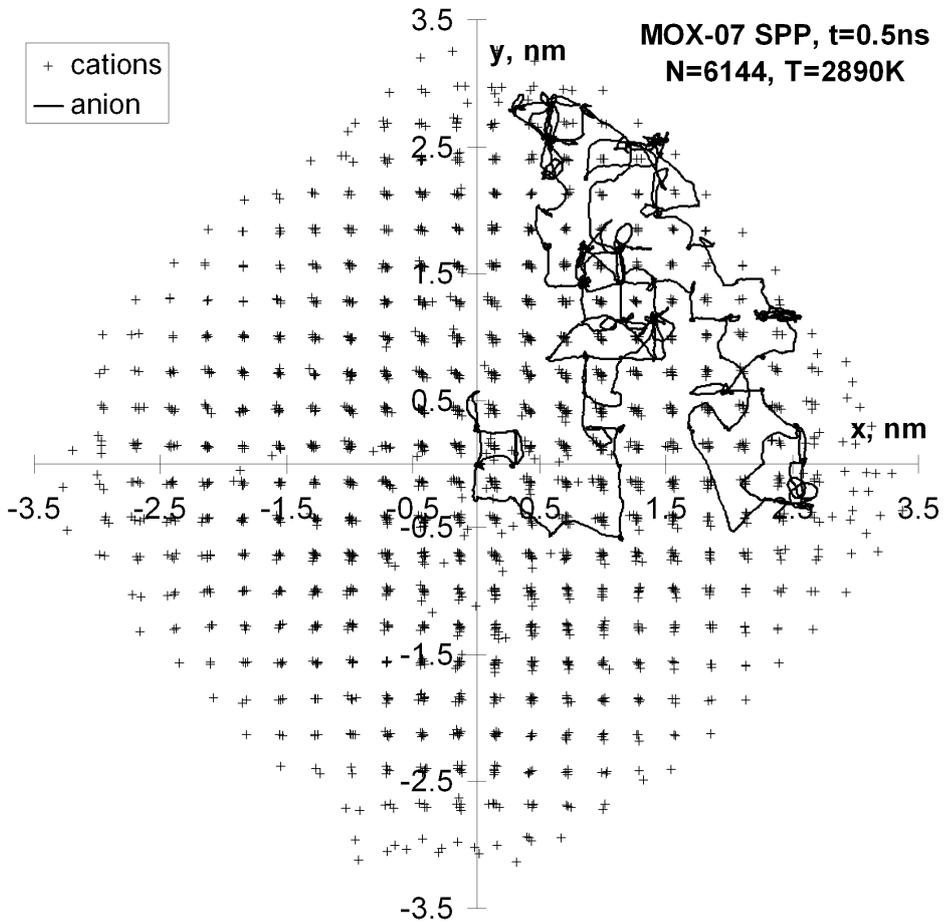

FIG. 7. Anionic diffusion in superionic phase with ordered surface.

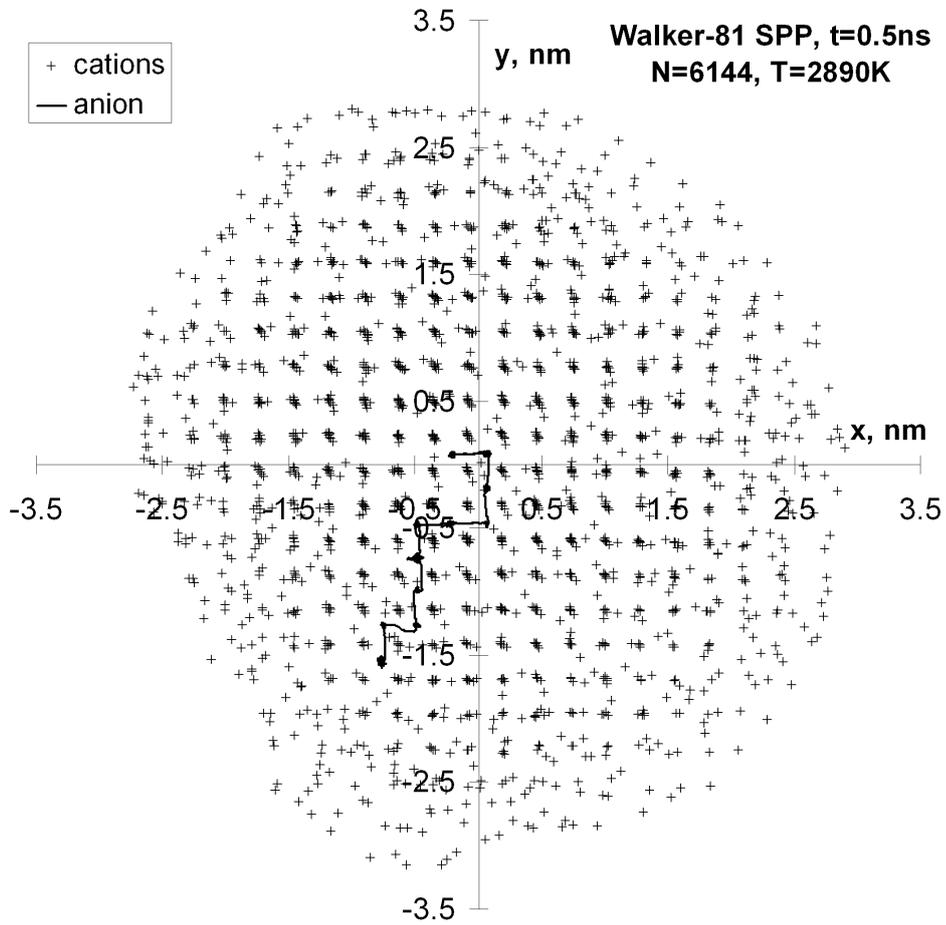

FIG. 8. Anionic diffusion in crystalline phase with disordered surface.

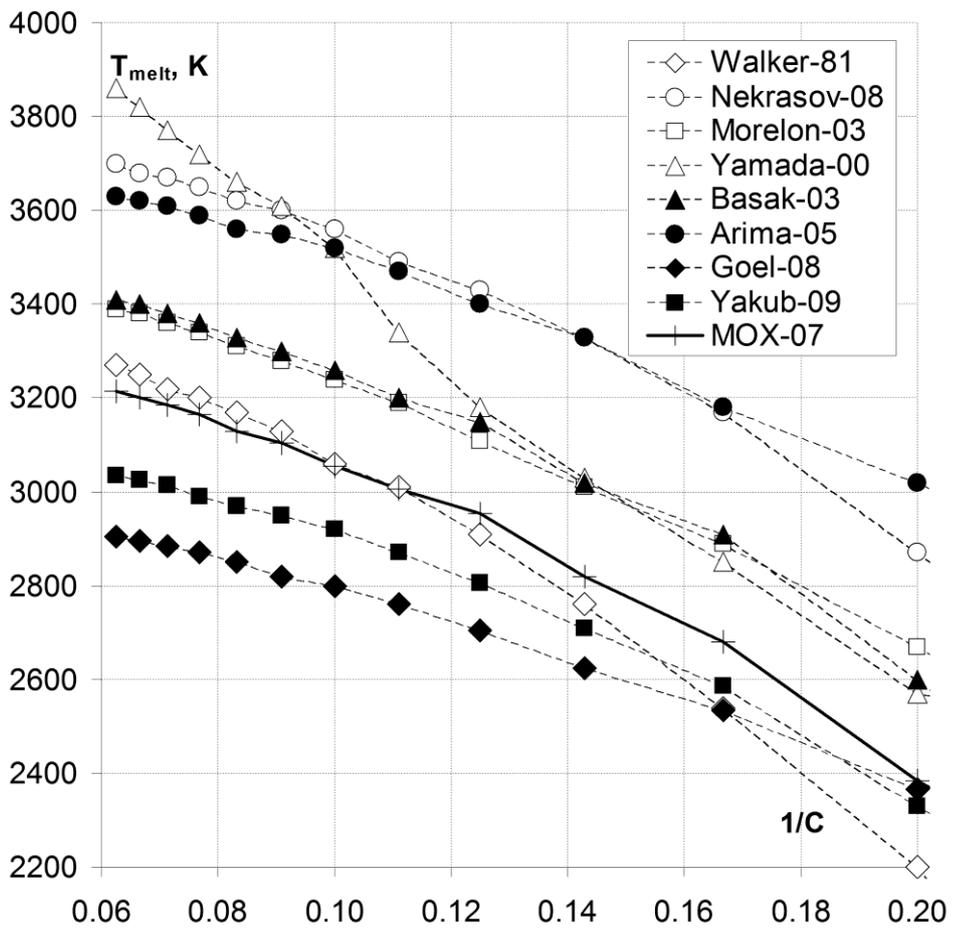

FIG. 9. Reciprocal size dependence of the nanocrystal melting temperature.

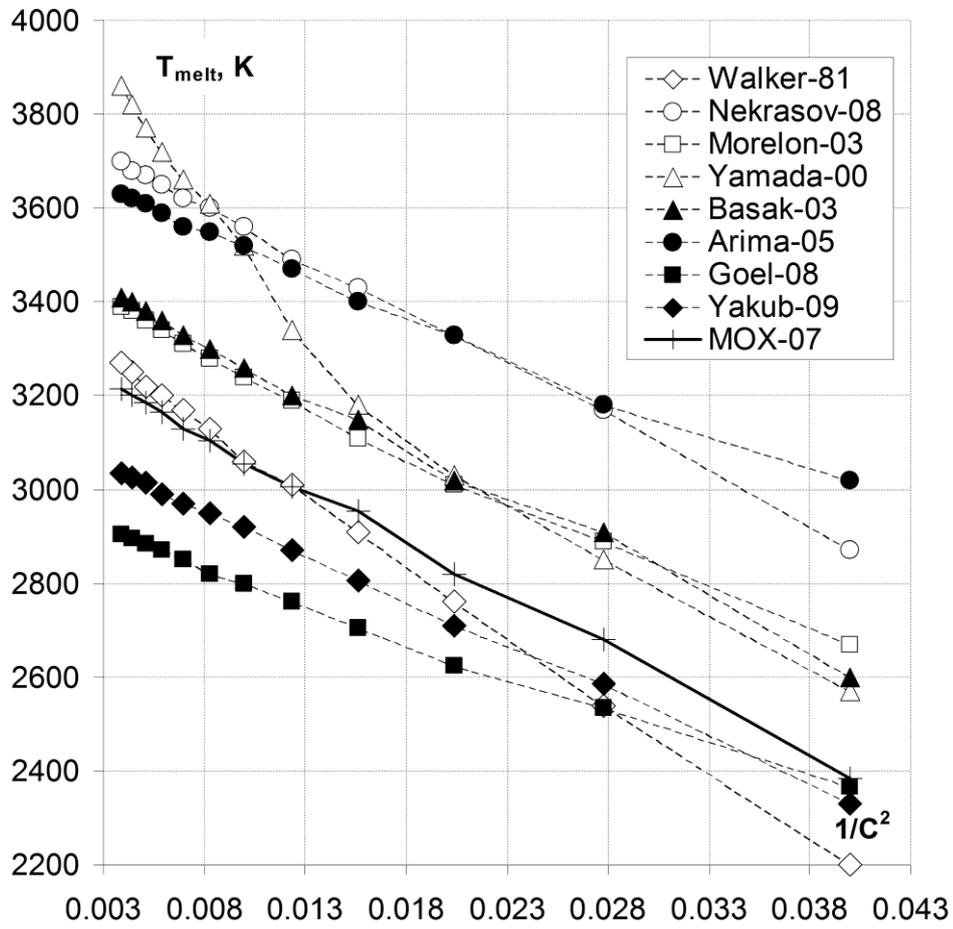

FIG. 10. Squared reciprocal size dependence of the nanocrystal melting temperature.

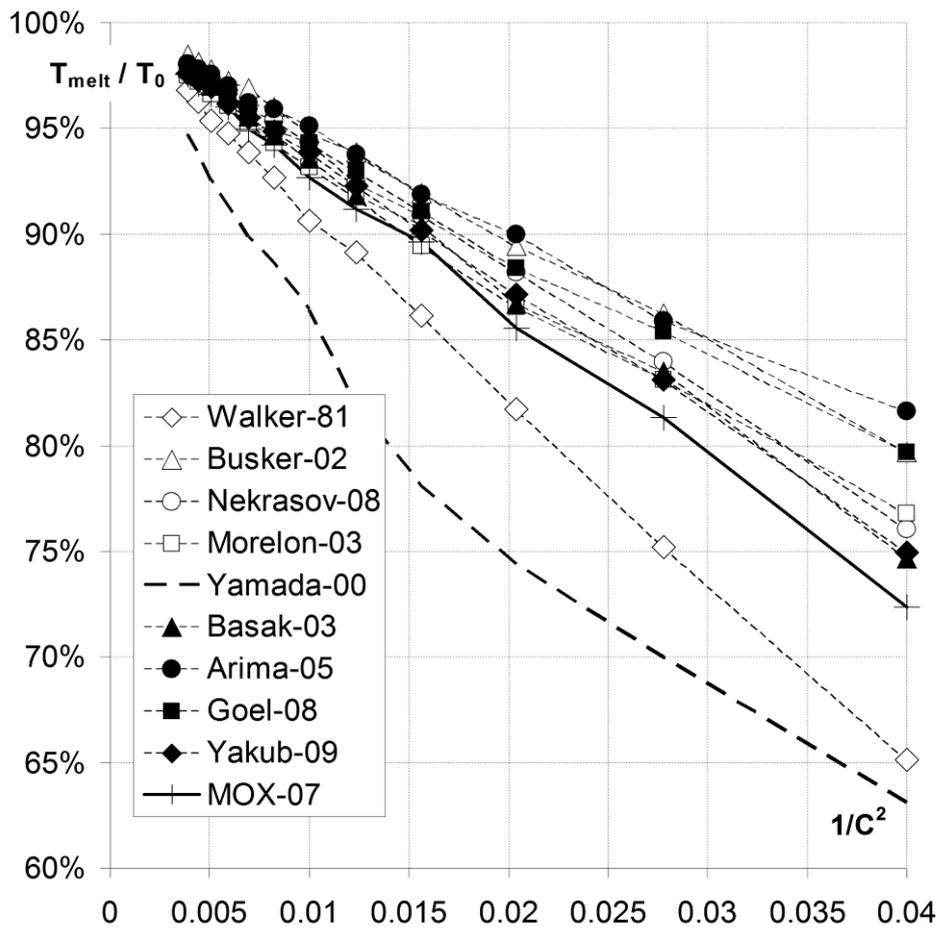

FIG. 11. Squared reciprocal size dependence of the relative melting temperature.

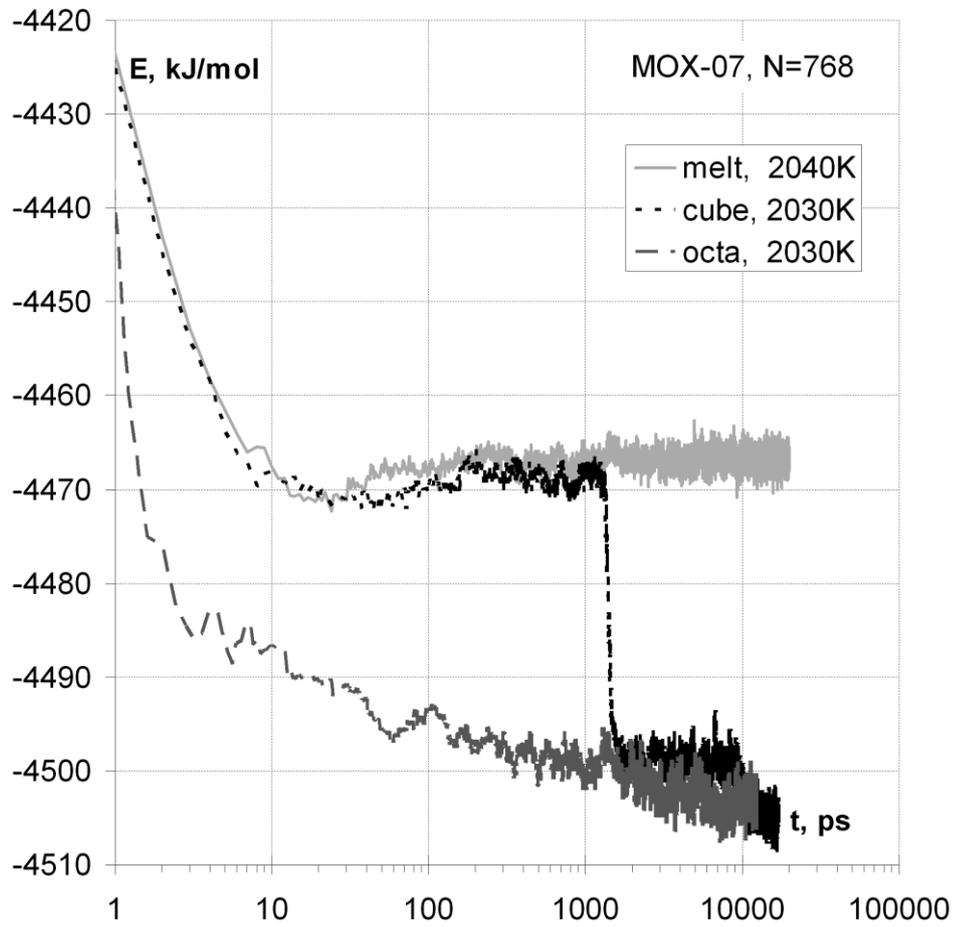

FIG. 12. Evolution of the energy of solid and liquid phases during small nanocrystal crystallization.

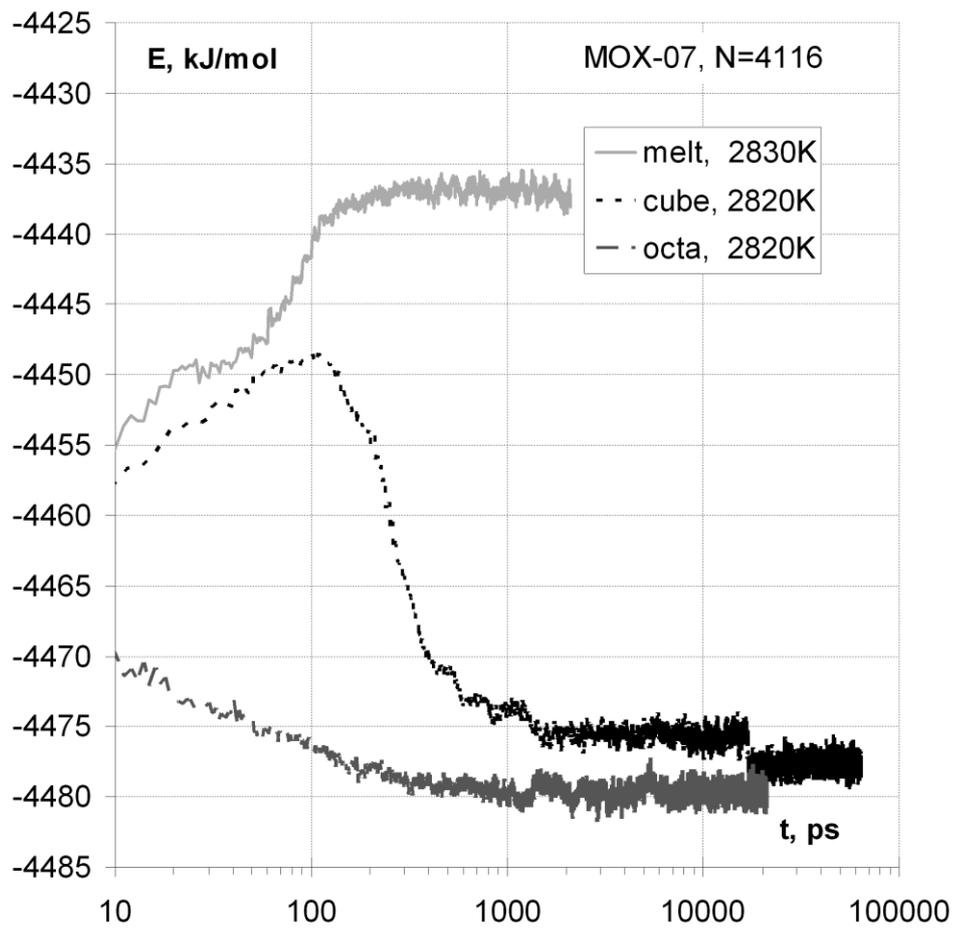

FIG. 13. Evolution of the energy of solid and liquid phases during large nanocrystal relaxation.

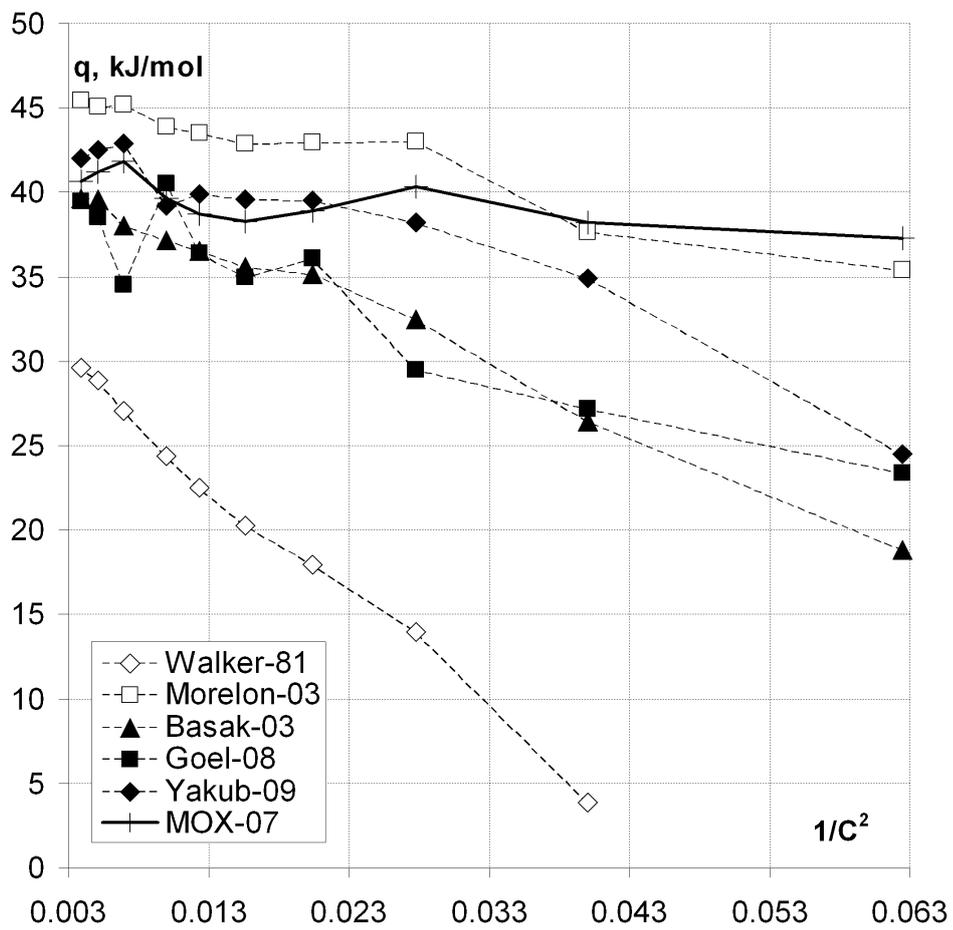

FIG. 14. Squared reciprocal size dependence of the nanocrystal heat of fusion.

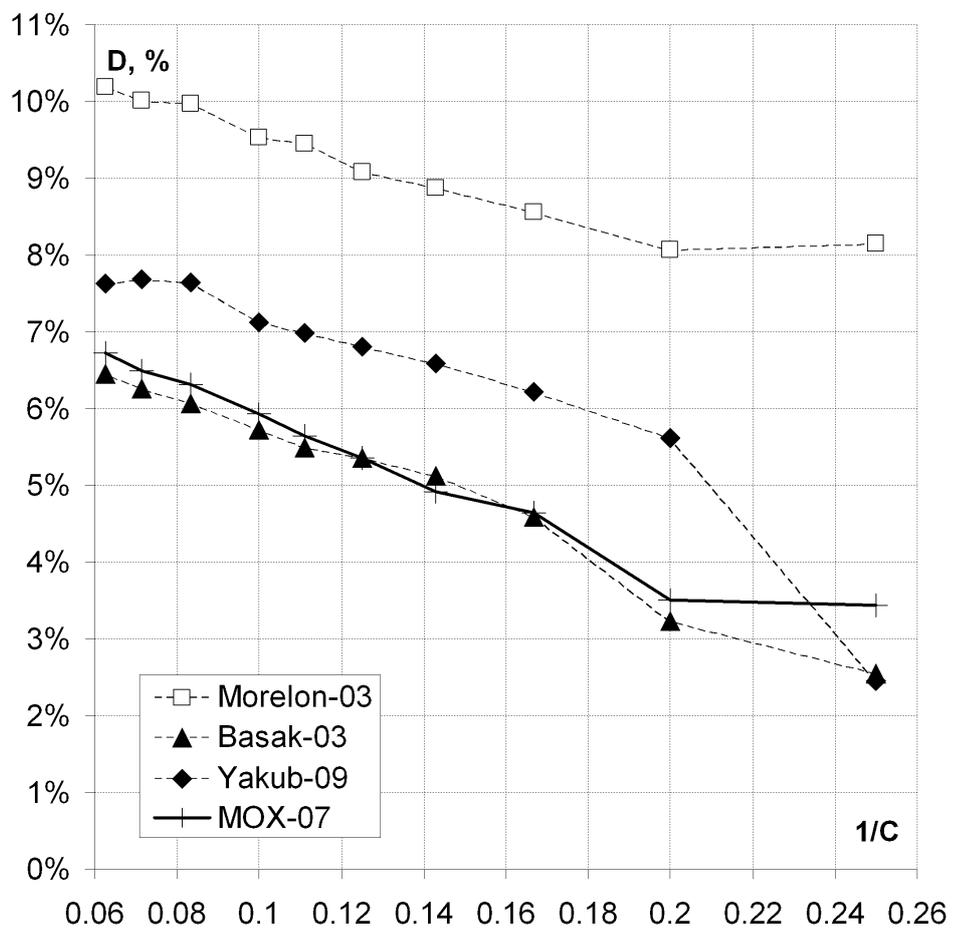

FIG. 15. Reciprocal size dependence of the nanocrystal density jump at melting.

TABLE 1. Dependence of the melting temperature on system size obtained under PBC.

| SPP | Q | PBC Melting temperature, K | | | |
|---|---|---|---|---|---|
| | | N=324 | N=768 | N=1500 | * N=12 000 |
| Walker-81 | 1 | 4900 | 4990 | 4980 | 5000 |
| Busker-02 | 1 | 6950 | 7110 | 7100 | 7100 |
| Nekrasov-08 | 0.95425 | 4950 | 5050 | 5030 | 5040 |
| Morelon-03 | 0.806813 | 4270 | 4260 | 4270 | 4260 |
| Yamada-00 | 0.6 | 4960 | 5000 | 5010 | 5000 |
| Basak-03 | 0.6 | 4170 | 4200 | 4200 | 4200 |
| Arima-05 | 0.675 | 4520 | 4550 | 4550 | 4550 |
| Goel-08 | 0.725 | 3840 | 3830 | 3840 | 3840 |
| Yakub-09 | 0.5552 | 3720 | 3760 | 3750 | 3750 |
| MOX-07 | 0.68623 | 4000 | 3990 | 4010 | 4000 |
| Recommendations | | ** $3140 \pm 20$ [21] [30] | | | |
| Experiment | | ** $3147 \pm 20$ [2] | | | |

* – measurements with step of 10 K instead of 1 K; ** – measurements in inert atmosphere.

TABLE 2. The density jump, the heat of fusion, the values of right and left parts of the Clausius-Clapeyron relation obtained under PBC.

| SPP | $1-V_{solid}/V_{melt}$ % | $V_{melt} - V_{solid}$, $10^{-6}$ m$^3$/kg | $H_{melt} - H_{solid}$ kJ/mol | dT/dP 0.01 K/MPa | $T_{melt}*\Delta V/\Delta H$ 0.01 K/MPa |
|---|---|---|---|---|---|
| Walker-81 | 24.4% | 32.3 | 46.5 | 23.8 | 93.6 |
| Busker-02 | 16.4% | 21.8 | 90.8 | 22.3 | 46.2 |
| Nekrasov-08 | 16.5% | 21.5 | 55.2 | 24.0 | 53.1 |
| Morelon-03 | 11.9% | 14.7 | 47.4 | 21.6 | 35.8 |
| Yamada-00 | 2.5% | 2.69 | 38.8 | 8.6 | 9.4 |
| Basak-03 | 8.4% | 9.92 | 53.8 | 15.6 | 20.9 |
| Arima-05 | 16.9% | 23.2 | 70.1 | 20.7 | 40.7 |
| Goel-08 | 20.1% | 27.8 | 52.5 | 22.9 | 54.7 |
| Yakub-09 | 10.3% | 12.5 | 52.0 | 16.2 | 24.4 |
| MOX-07 | 9.1% | 10.9 | 44.8 | 18.4 | 26.1 |
| Recommendations | 7.3% [30] | $8.26 \pm 3.30$ [2] [30] | $70 \pm 4$ [30] | – | 10.0 [30] |
| Experiments | 9.6% [31] | * $8.18 \pm 1.50$ [2] | $75 \pm 1$ [32] | $9.29 \pm 1.7$ [2] | – |

* – calculated by Manara et al. [2] using Clausius-Clapeyron relation from the experimental data [2] [32].

TABLE 3. Size dependence of the nanocrystal melting temperatures obtained under IBC.

| System size, ions | Walker-81 | Busker-02 | Nekrasov-08 | Morelon-03 | Yamada-00 | Basak-03 | Arima-05 | Goel-08 | Yakub-09 | MOX-07 |
|---|---|---|---|---|---|---|---|---|---|---|
| 768 | 1500 | 3960 | 2430 | 2320 | 2460 | 2140 | 2750 | 2110 | 1990 | 2040 |
| 1500 | 2200 | 4490 | 2870 | 2670 | 2570 | 2600 | 3020 | 2365 | 2330 | 2385 |
| 2592 | 2540 | 4860 | 3170 | 2890 | 2850 | 2910 | 3180 | 2535 | 2585 | 2680 |
| 4116 | 2760 | 5040 | 3330 | 3010 | 3030 | 3020 | 3330 | 2625 | 2710 | 2820 |
| 6144 | 2910 | 5180 | 3430 | 3110 | 3180 | 3150 | 3400 | 2705 | 2805 | 2955 |
| 8748 | 3010 | 5290 | 3490 | 3190 | 3340 | 3200 | 3470 | 2760 | 2870 | 3005 |
| 12 000 | 3060 | 5340 | 3560 | 3240 | 3520 | 3260 | 3520 | 2800 | 2920 | 3050 |
| 15 972 | 3130 | 5410 | 3600 | 3280 | 3610 | 3300 | 3550 | 2820 | 2950 | 3105 |
| 20 736 | 3170 | 5460 | 3620 | 3310 | 3660 | 3330 | 3560 | 2850 | 2970 | 3130 |
| 26 364 | 3200 | 5480 | 3650 | 3340 | 3720 | 3360 | 3590 | 2870 | 2990 | 3165 |
| 32 928 | 3220 | 5510 | 3670 | 3360 | 3770 | 3380 | 3610 | 2885 | 3015 | 3185 |
| 40 500 | 3250 | 5530 | 3680 | 3380 | 3820 | 3400 | 3620 | 2895 | 3025 | 3200 |
| 49 152 | 3270 | 5550 | 3700 | 3390 | 3860 | 3410 | 3630 | 2905 | 3035 | 3215 |
| Linear, T(1/C) | 3580 ± 26 | 5851 ± 13 | 3905 ± 22 | 3614 ± 16 | 4471 ± 14 | 3634 ± 16 | 3802 ± 18 | 3055 ± 6 | 3228 ± 29 | 3430 ± 5 |
| Parabolic, T(1/C$^2$) | 3377 ± 4 | 5634 ± 11 | 3775 ± 5 | 3477 ± 3 | 4072 ± 13 | 3484 ± 6 | 3701 ± 4 | 2969 ± 2 | 3110 ± 3 | 3296 ± 5 |
| PBC N=1500 | 4980 | 7100 | 5030 | 4270 | 5010 | 4200 | 4550 | 3840 | 3750 | 4010 |

TABLE 4. The calculated melting characteristics of a macrocrystal (i.e. extrapolated to infinite size).

| SPP | Boundary Conditions | T(X/Reff), K | X=2σv/q, nm | T(1/Reff$^2$), K | D(1/Reff), % | V$_{melt}$ −V$_{solid}$, 10$^{-6}$ m$^3$/kg | q(1/Reff$^2$), kJ/mol | T$_{melt}$*ΔV/ΔH, 0.01 K/MPa |
|---|---|---|---|---|---|---|---|---|
| Walker-81 | Periodic | – | – | 4980 | * 24.4 (27.8) | 32.3 | * 46.6 (66.5) | 93.6 |
| | Isolated | 3572 ± 32 | 0.463 ± 0.043 | 3376 ± 4 | 27.6 ± 0.4 | 34.9 | 31.8 ± 0.5 | 100.1 |
| Morelon-03 | Periodic | – | – | 4270 | * 11.9 (11.1) | 14.7 | * 47 (57.9) | 35.8 |
| | Isolated | 3608 ± 19 | 0.338 ± 0.026 | 3476 ± 4 | 11.2 ± 0.1 | 11.9 | 46.0 ± 0.4 | 24.3 |
| Basak-03 | Periodic | – | – | 4200 | * 8.4 (7.3) | 9.92 | * 54 (56.7) | 20.9 |
| | Isolated | 3627 ± 19 | 0.336 ± 0.026 | 3477 ± 7 | 7.47 ± 0.07 | 7.15 | 41.0 ± 0.3 | 16.4 |
| Goel-08 | Periodic | – | – | 3840 | * 20.0 (19.0) | 27.8 | * 51.2 (65.9) | 54.7 |
| | Isolated | 3051 ± 4 | 0.271 ± 0.006 | 2969 ± 2 | 19.7 ± 0.1 | 25.6 | 40.4 ± 1.3 | 50.9 |
| Yakub-09 | Periodic | – | – | 3750 | * 10.3 (8.7) | 12.5 | * 51.8 (57.2) | 24.4 |
| | Isolated | 3223 ± 26 | 0.327 ± 0.040 | 3105 ± 3 | 8.66 ± 0.12 | 8.93 | 42.9 ± 0.6 | 17.4 |
| MOX-07 | Periodic | – | – | 4010 | * 9.1 (8.0) | 10.9 | * 44.3 (51.1) | 26.1 |
| | Isolated | 3423 ± 1 | 0.345 ± 0.001 | 3291 ± 5 | 7.97 ± 0.09 | 7.59 | 40.7 ± 0.5 | 16.6 |

* – values outside the parentheses were obtained at the melting temperature under PBC; values in the parentheses were obtained at the melting temperature under IBC.